# Quantitative X-Ray Phase-Contrast Microtomography from a Compact Laser Driven Betatron Source

J. Wenz[1,2], S. Schleede[3], K. Khrennikov[1,2], M. Bech[3,4], P. Thibault[3], M. Heigoldt[1,2], F. Pfeiffer[3] and S. Karsch[1,2*]

[1] *Ludwig-Maximilians-Universität München, Am Coulombwall 1, 85748 Garching, Germany*

[2] *MPI für Quantenoptik, Hans-Kopfermann-Str. 1, 85748 Garching, Germany*

[3] *Lehrstuhl für Biomedizinische Physik, Physik-Department & Institut für Medizintechnik, Technische Universität München, 85748 Garching, Germany*

[4] *Medical radiation Physics, Clinical Sciences, Lund University, Barngatan 2:1, 22185 Lund, Sweden*

*\* stefan.karsch@mpq.mpg.de*

**X-ray phase-contrast imaging has recently led to a revolution in resolving power and tissue contrast in biomedical imaging, microscopy and materials science. The necessary high spatial coherence is currently provided by either large-scale synchrotron facilities[1] with limited beamtime access or by microfocus X-ray tubes[2] with rather limited flux.**

**X-rays radiated by relativistic electrons driven by well-controlled high-power lasers[3] offer a promising route to a proliferation of this powerful imaging technology. A laser-driven plasma wave accelerates and wiggles electrons, giving rise to brilliant keV X-ray emission. This so-called Betatron radiation is emitted in a collimated beam with excellent spatial coherence[4,5,6] and remarkable spectral stability. Here we present the first phase-contrast micro-tomogram revealing quantitative electron density values of a biological sample using betatron X-rays, and a comprehensive source characterization. Our results suggest that laser-based X-ray technology offers the potential for filling the large performance gap between synchrotron- and current X-ray tube-based sources.**

Since the discovery of X-ray radiation and its powerful imaging capabilities by Röntgen, X-rays have become a part of our daily life in medicine, industry, and research. The conventional technique of absorption imaging utilizes the large absorption variations of X-rays in matter (i.e. the imaginary part *β* of the refractive index *n=1-δ+iβ*) of different thickness and composition, limiting its usefulness to structures with high absorption gradients. However, the real part of the refractive index *δ* leads to phase variations of the X-rays depending on the sample's electron density. The latter can be explored with much higher sensitivity by phase-contrast (PC) techniques[7,8,9,10], which are far superior to conventional radiography for detecting structures in soft tissue with its rather homogeneous absorption profile. It is thus ideally suited for 3D investigations of (pathologic) tissue biopsies in medical research and diagnostics[9] but also finds applications in materials science. Computed tomography using PC images taken from different perspectives can provide the full 3D structure of the object with high resolution and enhanced contrast. PC imaging is complementary to coherent diffractive imaging[11], which is employed for tomographic reconstruction in the far field.

PC imaging can be implemented by free-space propagation[2], crystal analyzer-based[10], and crystal[9,12] or grating[8] interferometer-based techniques. For microscopy applications with micron-scale resolution, propagation-based PC imaging is the method of choice. It relies neither on additional X-ray optics, nor on substantial temporal coherence of the source. The only requirement is that the transverse coherence length, given by $l_t = \lambda \cdot R/\sigma$ is larger than $\sqrt{\lambda D}/(2\sqrt{2}) \cong O(1-10 \mu m)$[13]. Here, $\lambda$ is the wavelength, $R$ is the source-detector distance, $\sigma$ is the source size and $D = d \times l/(d+l)$ the defocusing distance with $l$ as source-sample and $d$ sample-detector distances. This is met by third-generation synchrotrons, but their size and cost prevent their proliferation in hospitals and research institutions. Microfocus X-ray tubes provide the desired spatial coherence but suffer from a modest X-ray flux, implying lengthy exposure times. In order to spark off widespread application of this powerful technique, a compact high-brilliance source is needed.

Laser-wakefield acceleration of electrons[3] (LWFA) (see methods) has already been studied as a possible future alternative. LWFA-electron beams exhibit transverse emittances comparable to the best conventional linear accelerators and bunch durations of approximately 5 fs FWHM[14,15], rendering them unique among compact sources. During the acceleration process the electrons are also wiggled transversely by the strong radial fields of the plasma wave, causing them to emit a forward-directed, incoherent X-ray beam, referred to as betatron radiation (see methods).

Recent research demonstrated the potential of laser-driven betatron sources for recording single-shot X-ray PC images based on the free-space propagation technique[6]. Due to advances in LWFA stability by using a high contrast driver pulses and turbulent free steady-state gas flow target[16,17] and the photon flux of the betatron source, we were for the first time able to record and quantitatively reconstruct a phase tomogram of a complex object. As an illustrative example, we chose a dried insect (Chrysoperia carnea, green lacewing). We achieved this by evaluating a set of 1487 single-shot PC images taken from various angles. The micro-radiography images featured a field of view of 7.5x6.9 mm$^2$ and a resolution of 6 μm, limited by the CCD area and pixel size, respectively. For quantitative reconstruction of the tomogram[18] from these images, we performed a careful characterization of the X-ray source in terms of its transverse dimensions and spectrum.

In the experiment, intense short laser pulses were focused at the entrance of a hydrogen-filled gas cell (see Fig. 1 and methods section). For our pulse parameters the highest electron energies with a peak at 400 MeV, lowest divergences 1.3±0.2 mrad FWHM and moderate charge (50 pC) were produced at plasma electron densities of $5\times10^{18}$ cm$^{-3}$, created from laser-induced field ionization of the target gas. The betatron motion leads to the emission of well-collimated X-ray beams with a spectrum peaked at 4.9 keV, measured by their transmission through a stack of filters[19]. However, when the plasma density is increased to $1.1\times10^{19}$ cm$^{-3}$, the electron energy drops to 200 MeV and their divergence increases up to 5 mrad, along with a substantial increase of electron beam charge to 400 pC. Now the X-ray beam divergence triples from 2,3±0.2 mrad to 6,0±1.1 mrad and the photon fluence increases by more than an order of magnitude. The energy spectrum stays roughly constant, partly because higher wiggling fields in the dense plasma offset the lower electron energies. The increase in photon number results from the dramatic increase of trapping efficiency at higher densities. It is evident from Fig. 2a that despite large shot-to-shot variations in the electron spectrum in this high-density regime, the X-ray spectrum is remarkably stable, a behavior not reported before. We attribute this to the fixed low-energy cutoff of the aluminum laser blocking filter (see Fig. 1 and methods) on the one hand and to the incoherent superposition of emission from many electrons with a broad spectrum on the other hand. This betatron-optimized regime results in the emission of $1.2\times10^9$ photons/msr/shot (±20%) above 1 keV, at the position of the object.

The source size was derived from the Fresnel diffraction pattern of an object in the X-ray beam, as shown in Fig. 2b. Comparison of a modeled diffraction pattern with the data yields a source size of 1.8±0.1 µm rms. Assuming pulse durations of 5 fs as suggested by numerical studies, recent reports[14,15] and our own yet unpublished measurements, the source exhibits a peak brilliance of $2\times10^{22}$ photons/(s mm$^2$ mrad$^2$ 0.1% bandwidth) at the position of our sample. In the current proof-of-principle experiment, the average brilliance and photon flux density at the sample was limited to $1\times10^7$ photons/(s mm$^2$ mrad$^2$ 0.1% bandwidth) and $2\times10^7$ photons/(s cm$^2$), respectively, owing to a shot rate of 0.1 Hz due to gas load in the chamber and data acquisition limitations. An optimized pumping design and improved data acquisition permitting the full 5 Hz short rate of the laser would improve these figures by a factor of 50 and yield few-minute scan times.

The scalability of the photon energy depends on the electron energy, the plasma density and the wiggler strength parameter, which in LWFA are all interlinked[3]. Clever target engineering, i.e. separating acceleration and radiation zone, off-axis injection[20] or laser-betatron resonance effects[21] may strongly enhance the betatron amplitude and hence the critical energy. In [21], a 20-keV X-ray spectrum with a tail to 1 MeV was achieved with laser pulses containing only three times more energy. The shot-to-shot stability of our X-ray source is excellent for a laser-driven process, yielding more than $10^7$ ph/shot in >99% of the laser shots in the tomographic scan, with low fluctuations of the X-ray spectrum (see Fig.2a), and a photon number constant to within ±20% rms, making it suitable for multi-exposure tomography.

For the tomography scan the object was placed on a rotating mount into the X-ray beam (see Fig. 1). Although a dried sample exhibits different electron densities compared to a fresh tissue, it was chosen for technical simplicity for this proof-of-concept experiment. Its ability to be put within the vacuum chamber avoided unnecessary intermediate air propagation (as the cooled CCD chip needs to be evacuated), and consequently unnecessary transmission losses in X-Ray windows.

The raw PC images recorded on the CCD exhibit the so-called edge-enhancement effect (Fig. 3a), which is inherent to propagation-based PC imaging[13] in the Fresnel-diffraction regime. No optical elements between the source and the detector are used, but the wave propagates sufficiently far beyond the sample (1.99 m) for Fresnel diffraction to occur. The edge-enhanced image is useful by itself for visual inspection when high-resolution features with poor absorption contrast (Fig. 3a) are of particular interest. However, propagation-induced intensity fringes of a pure phase object are not a direct measure of the phase shift but rather the Laplacian of the phase front[2]. A reconstruction of raw phase projections will thus only yield gray level variations at material interfaces (Fig. 4a). Due to the missing link of reconstructed contrast to material properties, a quantitative analysis and automatic segmentation via thresholding is not possible.

In absorption tomography, projections of the linear absorption coefficient along the beam are directly obtained from the logarithm of the recorded intensity. The subsequent reconstruction exhibits area contrast with gray values directly related to material properties of the sample under investigation. Starting from diffracted intensity measurements at a certain propagation distance, phase-retrieval algorithms are employed to create line projection images of the refractive index decrement $\delta$ (phase maps, see methods). The transport-of-intensity-equation (TIE) relates the edge-enhanced image measured at the detector to the phase distribution at the exit plane of the sample. As we only used one propagation distance we employed a single-material constraint to solve the TIE and retrieve phase maps of the insect (Fig 3b) (see methods).

As the retrieved phase map is directly related to the integrated decrement $\delta$ of the index of refraction, the reconstruction yields information on electron density distribution in the sample. Before reconstruction, the 360 projections taken over 360° were each averaged over four subsequent laser shots and binned by a factor of two in both image dimensions to yield an artifact-free reconstruction (see methods). Standard filtered back projection (FBP) was used to reconstruct the transverse slices shown in Fig. 4 b and c. The reconstruction reveals a distinct contrast between insect and background, allowing segmentation via simple thresholding. Additional TEM measurement of the insect's leg revealed that the reconstructed electron densities are in good agreement -under the consideration of the resolution of our setup- with the expected electron densities for chitin (see supplementary material).

A 3D rendering of the sample is presented (see Fig. 5 and Supplementary Movie), including sectioning planes of the 3D volume with gray levels corresponding to electron density. This result demonstrates that laser–driven X-ray sources have reached the verge of practical usefulness for application-driven research. If further progress regarding mean photon flux can be made, laser-driven sources due to their compactness, relatively low cost and high peak brilliance might become valuable tools for university-scale research. Especially in view of the ongoing dynamic evolution in high-energy, high-repetition rate laser technology[22,23] aiming to scale multi-TW lasers to kHz and beyond, average fluences approaching current state-of-the art compact synchrotron sources are expected to become available in the near future. This might help proliferation of biomedical X-ray diagnostics, such as tumor early detection, X-ray microscopy, and non-destructive industrial testing for static – as demonstrated here by a micro-tomogram of a biological sample– and in particular time-resolved imaging with femtosecond resolution on a laboratory scale.

## Methods

**Laser wakefield acceleration**[3]**.** The experiments were performed using the ATLAS Ti:Sapphire laser at the MPI for Quantum Optics. It occupies a table area of approx. 15m$^2$, and delivers 1.6 J-energy, 28 fs-duration (60 TW) laser pulses, centered at 800 nm wavelength. They are focused onto the entrance of a Hydrogen-filled gas cell by an off-axis parabolic mirror (f=1.5m, F/20) to a spot size of 22μm FWHM. This corresponds to a vacuum peak intensity of 1x10$^{19}$ W/cm$^2$. At these intensities, the hydrogen is fully ionized and the laser propagates through plasma. Its strong ponderomotive force drives a plasma wave with frequency $\omega_p = \sqrt{\frac{n_e e^2}{m_e \varepsilon_0}}$. Here, $n_e$, $e$, and $m_e$ are the density, charge and rest mass of the plasma electrons, and $\varepsilon_0$ is the vacuum permittivity. The wave phase velocity matches the laser group velocity, so that in the frame of the laser pulse, the plasma wave constitutes a co-moving accelerating field. Electrons from the plasma are trapped into this wave by wavebreaking and accelerated to relativistic energies around 200-400 MeV, depending on the electron density. A magnet deflects the electron beam according to energy onto a scintillating screen (see Fig. 1). From the position and brightness on the screen the energy, divergence and charge of the electron bunch can be deduced[24].

**Betatron radiation:** The plasma wavelength $\lambda_p = \frac{2\pi c}{\omega_p}$ confines the wakefield to a radius of approx. 10 μm, causing strong radial fields that force the electrons into anharmonic transverse betatron oscillations at a frequency $\omega_\beta = \frac{\omega_p}{\sqrt{2\gamma}}$ during acceleration. Here, $\gamma$ is the relativistic factor of the electron beam. The strong radial fields lead to large angular excursions, while keeping the beam size within a few microns, triggering the emission of high harmonics of $\omega_\beta$ in a co-moving frame. $\omega_\beta$ varies during acceleration, causing incoherent emission. In the laboratory frame, the radiation is relativistically boosted to the X-ray regime and merges into a synchrotron-like continuum, described by a critical energy $E_{crit} = 3\frac{h}{2\pi}K\gamma^2\omega_\beta$. Here, K is the wiggler parameter defined through $K \approx \gamma\theta$[5], where $\theta$ is the opening angle of the emitted radiation. For our experimental conditions, even a small off-axis distance of 1 μm corresponds to a wiggler parameter on the order of 10, leading to high harmonic orders.

For the spectrum and source size measurements, the x-rays are freely propagating from the source to an Andor model DO432 BN-DD back-illuminated CCD detector at a distance of 3.26 m. The on-axis laser light is blocked behind the source by two 10 μm thick Al-foils which are transparent for radiation above 1 keV. A filter cake with different material thicknesses for spectrum characterization and a Tungsten wire for source size analysis can be moved into the beam. For the tomography studies the sample is mounted $l$=0.73 m from the source, $d$=1.99 m in front of the detector, yielding a 3.7x magnification.

**Modeling of Fresnel diffraction.** The source size was derived by analyzing the Fresnel diffraction pattern from a tungsten wire backlit by the X-ray beam. The measured edge diffraction on the detector from a 100 μm thick Tungsten wire (26 cm behind the source) is shown in the inset of Fig. 2b, and was compared to modeled distributions for various source sizes. They were obtained by summing up the

Fresnel diffraction from a knife edge, as described in e.g. Born & Wolf[25] for all energy bins of the incident spectrum weighted by the CCD response. The beam, showing a Gaussian shape on the CCD chip, was assumed to be Gaussian at the origin. The result for different source sizes is shown in the inset of Fig. 2b. The information about the source size is completely indicated by the first overshot of the profile. In order to improve the signal to noise ratio we have vertically summed up the profiles, taking into account the curvature of the wire by a cross-correlation between different rows. Our analysis yields a best fit for a source size of 4.2±0.2 μm FWHM.

**Phase retrieval**: The transport-of-intensity equation (TIE) directly relates the phase distribution in the planes orthogonal to the optical axis to the propagation of the wave-front intensity of the beam. A variety of phase-retrieval algorithms, which solve the TIE, have been proposed, differing in raw data input needed (e.g. multiple sample to detector distances) and additional constraints on sample material properties[26]. We use a single-distance quantitative phase retrieval method that does not require the sample to show negligible absorption[27]. It employs a single material constraint corresponding to a fixed $\delta/\beta$ ratio representing the sample's main chemical component. If this assumption is justified and the sample exhibits comparably weak absorption such as in our case, this approach allows for a quantitative reconstruction of electron density values in the sample as shown in [28]. The projected thickness of the sample which is directly related to the phase shift imposed onto the wavefront via $\varphi(\vec{r}) = -2\pi/\lambda_{mean} \delta_{poly} T(\vec{r})$, can be retrieved by using the following equation[27]:

$$T(\vec{r}) = -\frac{1}{\mu_{poly}} \ln\left( F^{-1}\left\{ \frac{ \frac{F(I(M\vec{r}))}{I_0} }{ 1 + \frac{R\delta_{poly}}{M\mu_{poly}} |\vec{k}|^2 } \right\} \right) \quad , (1)$$

Here T is the retrieved thickness of the sample, $\vec{r}$ are the transverse coordinates, $\vec{k}$ are the Fourier space coordinates, $I$ is the measured intensity, $I_0$ is the uniform intensity of the incident radiation, $M$ is the magnification of the image, $R$ is the distance from the sample to the detector and $\mu_{poly}$ and $\delta_{poly}$ are the material-dependent linear absorption coefficient and refractive index decrement, respectively. In the case of a polychromatic X-ray spectrum the most accurate phase retrieval results are achieved through the calculation of effective $\mu$ and $\delta$ values. We assume the main chemical composition present in the dried insect to be Chitin ($C_8H_{13}NO_5$). Values of $\mu_{poly}$=70.15 cm$^{-1}$ and $\delta_{poly}$= 1.38x10$^{-5}$ were calculated from reconstructed X-ray spectra (see Fig. 2a), tabulated $\delta(E)$ and $\beta(E)$ values at a density of $\rho$=2.2 gcm$^{-3}$ [29], and the known detector response function[30]. The phase map as depicted in Fig. 3b) was reconstructed using Eq. (1) with values of $R$=199 cm, $M$=3.7 and a mean energy as seen by the detector of $E_{mean}$=8.8 keV ($\lambda_{mean}$=1.4 Å).

**Treatment of raw images.** The fluctuations of the X-ray point of origin (12 μm rms vertical, 18 μm rms horizontal) are caused by shot-to-shot laser pointing fluctuations. To correct for these, prior to reconstruction all images were be registered using normalized cross correlation. The shape of the correlation surface is assumed to fit two orthogonal parabolic curves. Sub-pixel registration accuracy is obtained by fitting a paraboloid to the 3x3 pixel vicinity of the maximum value of the cross correlation matrix. Sub-pixel shifting is performed in Fourier space. To account for the Gaussian intensity profile of the X-ray beam the sample is masked out using an edge detection filter and images are background corrected by subtracting a second order polynomial. The vertical alignment of the tomography scan is performed using cross correlation of integrated pixel values perpendicular to the tomography axis. Horizontal alignment was performed manually.

Supplementary information is linked to the online version of the paper at www.nature.com/nature.

# Acknowledgment

This work was supported by DFG through the MAP and TR-18 funding schemes, by EURATOM-IPP, and the Max-Planck-Society. S.S., M.B., and F.P. acknowledge financial support through the DFG Gottfried Willhelm Leibniz program and the European Research Council (ERC, FP7, StG 240142). S.S. acknowledges the Technische Universität München graduate School for the support of her studies. P.T. acknowledges support from the ERC starting grant "OptImaX" (#279753).

# Author contribution

J.W., S.S., K.K., M.H., F.P. and S.K. designed and carried out the experiments. J.W., S.S. and K.K. performed the source characterization. S.S., M.B. and P.T. analyzed the tomography data. J.W. and S.S. wrote the main part of the paper. F.P. and S.K. provided overall guidance and supervised the project. All authors discussed the results, reviewed and commented on the manuscript.

# Author information

Correspondence and request for materials should be addressed to S.K. (stefan.karsch@mpq.mpg.de).

# Figure Captions

**Figure 1 Experimental Setup:** The laser pulse (1.6J, 28fs, 60 TW) (red) is focused by a *F*/16 off-axis parabola to a 22 µm diameter spot size on the entrance of a 6 mm long gas cell with an plasma electron density of $n_p=1.1 \times 10^{19}$ cm$^{-3}$. The residual laser light is blocked by a 10 µm thick Aluminum foil in front of the detector. Dipole magnets deflect the accelerated electrons (yellow) away from the laser axis onto a scintillating screen (LANEX). The X-ray beam (pink) transmitted through the foil is recorded by a cooled back-illuminated X-ray CCD with 22.5 µm square pixels. The tomogram was acquired in the experimental configuration as shown. The thickness of the Al-foil in front of the sample was 20 µm in order to protect it from the laser light during the scan. The 'Al- cake' and wire-arrangement each protected by an Al foil of 10 µm thickness can be inserted into the X-ray beam for spectral and source size characterization. The Al foil of 10 µm thickness in front of the CCD was present during whole experiment

**Figure 2 Characterization of the Betatron source. a**: Electron and corresponding X-ray spectra as seen by the sample for 18 individual laser shots. The X-ray spectra were reconstructed from the transmission signal of the filter cake with overall thicknesses ranging from 20-630 µm (see left inset in Fig. 2b). Inside the red circle, corresponding to $(1.35 \times 10^{-2})$ msr, $(1.6 \pm 0.3) \times 10^7$ photons are detected and analyzed for their energy. Even for large electron energy fluctuations, the X-ray spectral shape is remarkably stable. **b**: Source size measurement: A comparison of the measured intensity distribution integrated along a 100 µm thick tungsten wire (right inset) and the modeled intensity distribution for a Gaussian source spot, taking into account the spectra in Fig. 2a, reveal a best value of 1.8 µm rms.

**Figure 3 Lacewing insect (chrysoperia carnea), imaged with Al-filtered betatron x-ray spectrum. a:** The image shows a selected single-shot radiograph dominated by X-ray inline phase contrast. Small details are highlighted due to the strong edge enhancement effect (see insets of magnified sections). **b:** The corresponding quantitative phase map was retrieved using a single-material constraint. A series of these maps is used for the reconstruction of the insect as shown in Figs. 4 and 5.

**Figure 4 Transverse slices of the sample. a:** A reconstructed transverse slice of the lacewing without phase retrieval (i.e using raw PC images as in Fig.3a) highlights material boundaries but does not allow for a quantitative analysis. **b:** The same transverse slice reconstructed after phase retrieval, using phase images as in Fig. 3b. **c:** Reconstructed transverse slices as in b with gray values representing electron density values. The reconstruction exhibits good area contrast allowing for volume rendering and segmentation as shown in Fig. 5. Electron density scale applies to subfigures b and c.

**Figure 5 Volume Rendering. a:** Photograph of the sample. **b:** 3D rendering of the sample, imaged with Al filtered betatron X-ray spectrum. **c,d:** Cutting planes of the 3D volume are shown with their gray levels corresponding to electron density distribution in the sample.

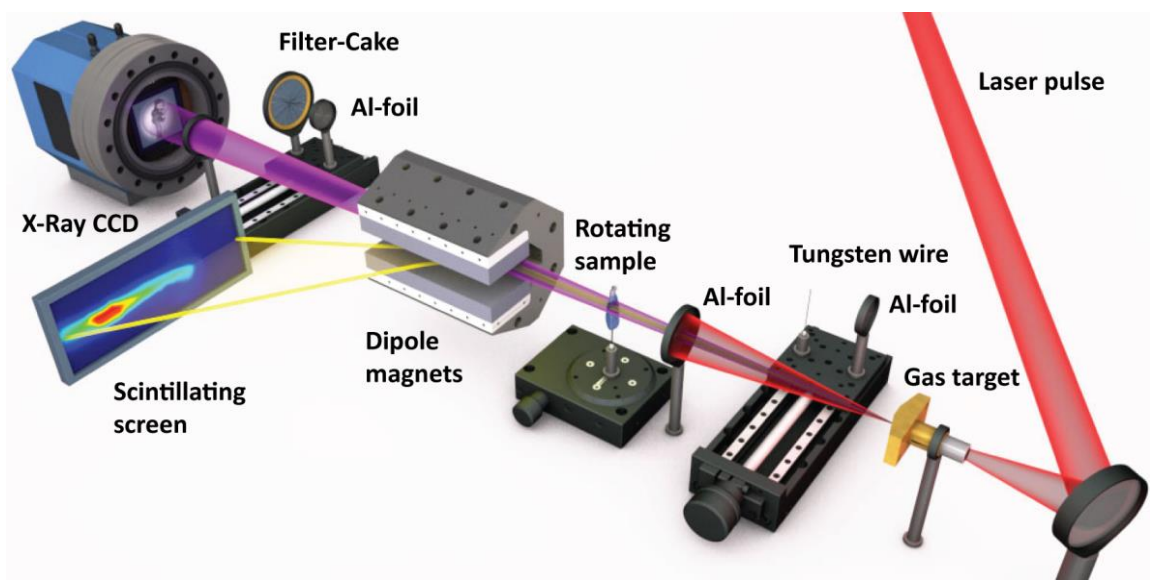

**Fig 1.**

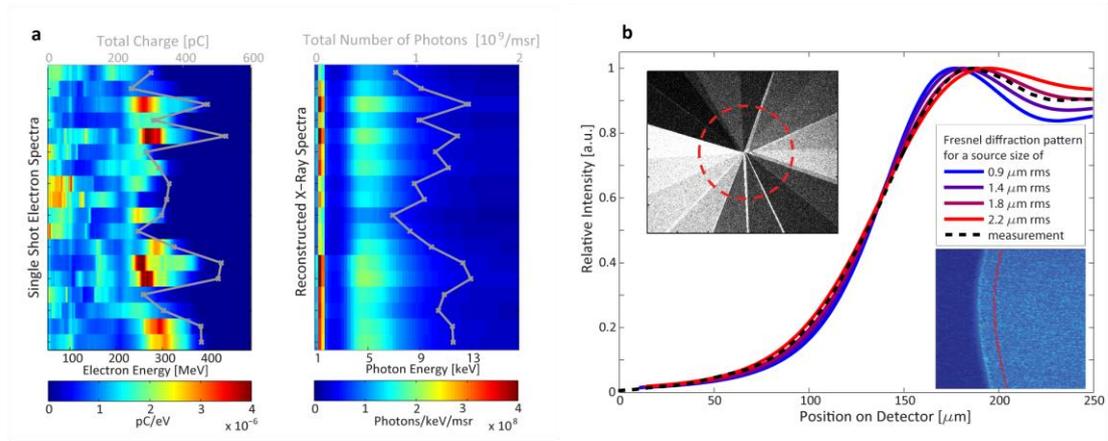

**Fig 2.**

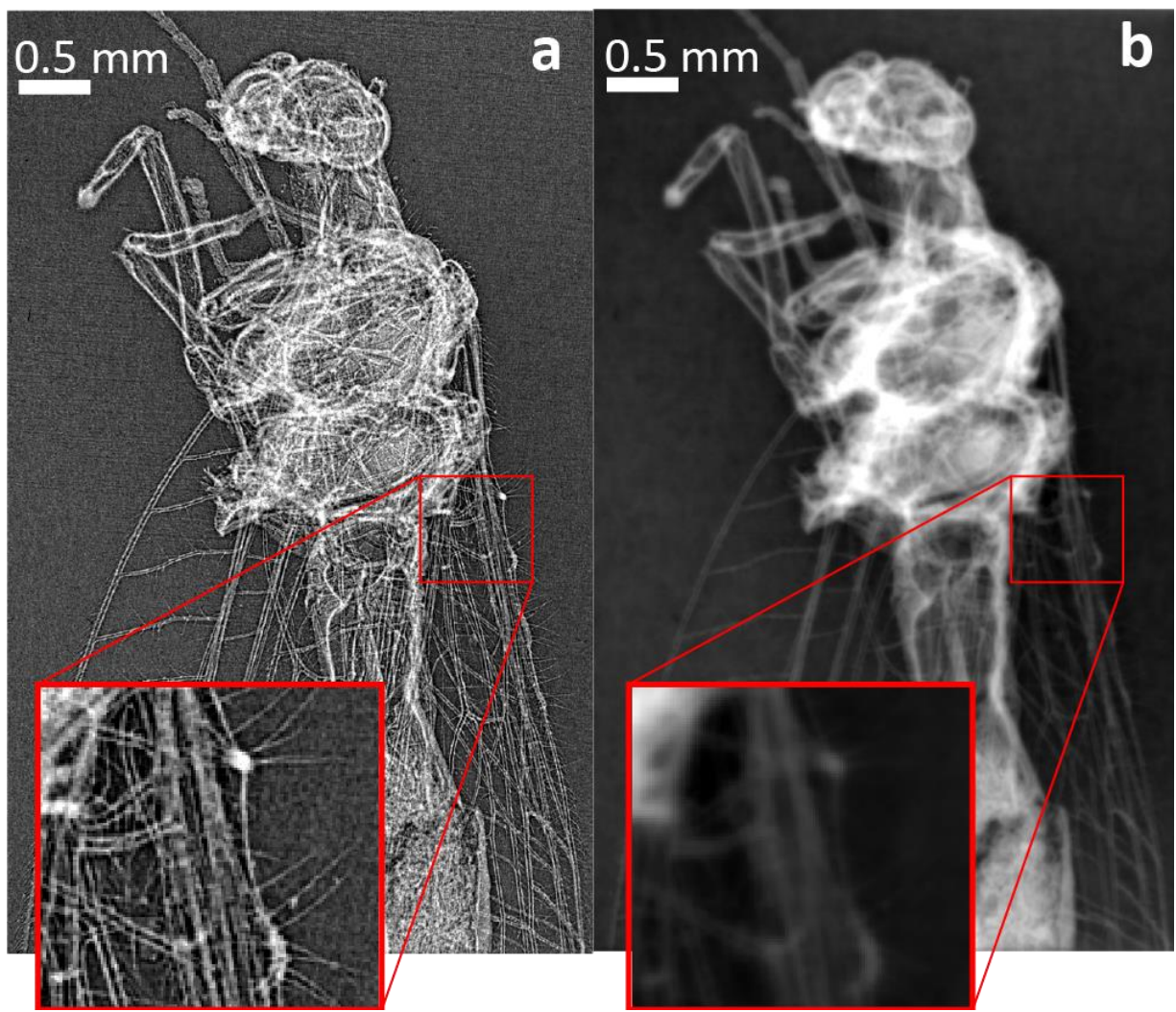

**Fig.3**

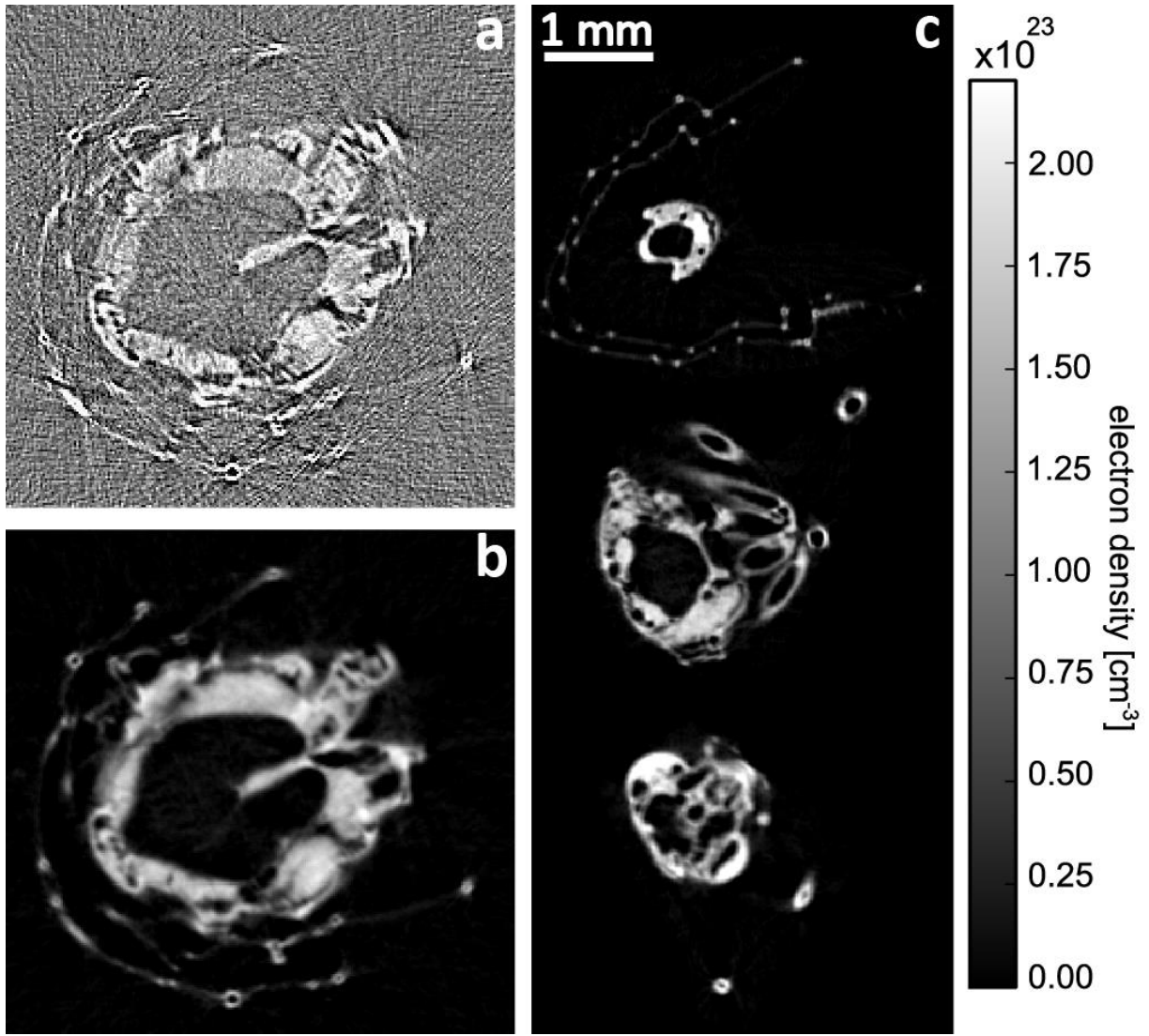

**Fig. 4**

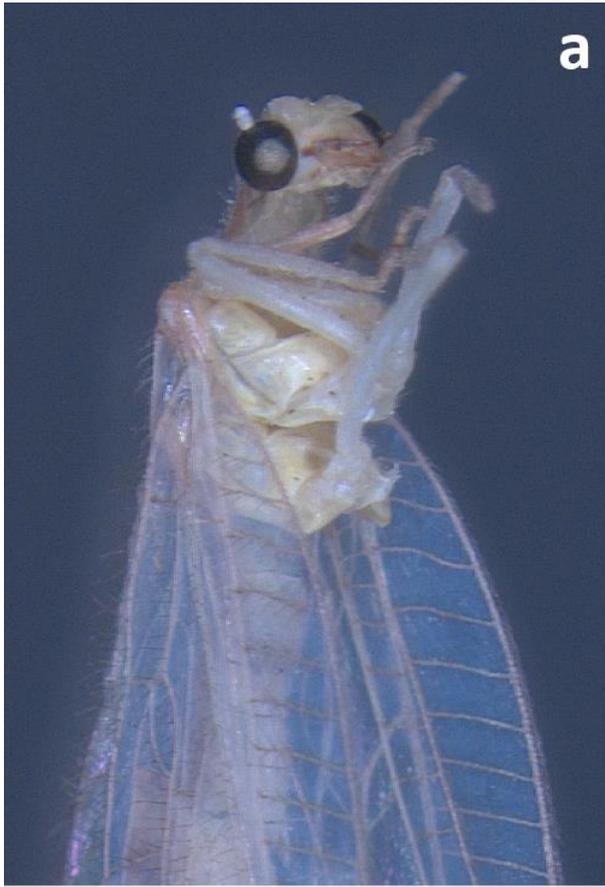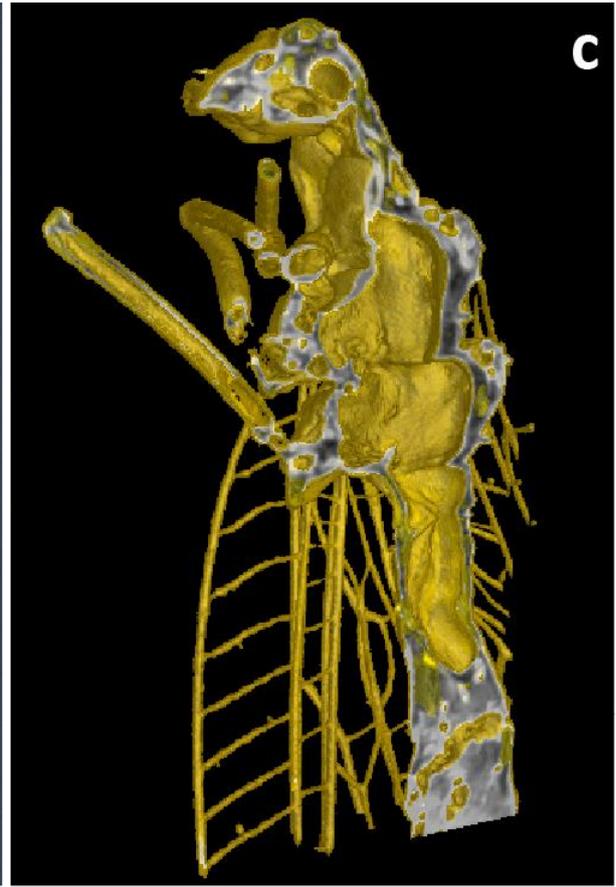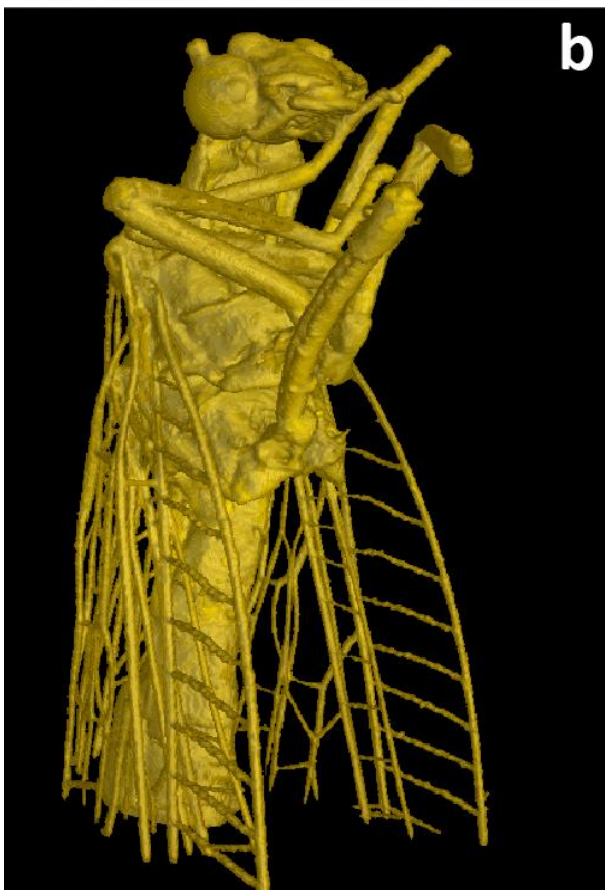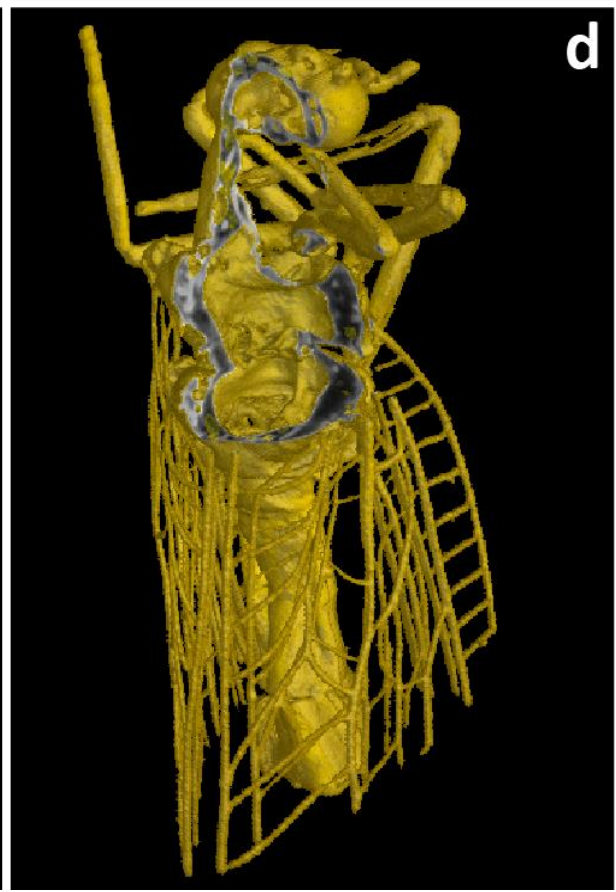

**Fig. 5**